\documentclass[a4paper,fleqn,usenatbib]{mnras}

\usepackage{newtxtext,newtxmath}

\usepackage[T1]{fontenc}
\usepackage{ae,aecompl}

\usepackage{graphicx}	
\usepackage{amsmath}	
\usepackage{amssymb}	
\usepackage{multicol} 
\usepackage{array}

\newcommand{\nscool}{\texttt{NSC\textsc{ool}}\,}
\newcommand{\kteff}{$kT^{\infty}_{\mathrm{eff}}$\,}
\newcommand{\nh}{$N_\mathrm{H}$\,}

\newcommand{\mdot}{$\langle \dot{\mathrm{M}} \rangle$\,}

\title[Further constraints on the crustal properties of 1RXS J1804]{Further constraints on neutron star crustal properties in the low-mass X-ray binary 1RXS J180408.9$-$342058}

\author[A. S. Parikh et al.]{
A. S. Parikh,$^{1}$\thanks{E-mail: a.s.parikh@uva.nl}
R. Wijnands,$^{1}$
N. Degenaar,$^{1}$
L. Ootes,$^{1}$ and
D. Page$^{2}$
\\
$^{1}$Anton Pannekoek Institute for Astronomy, University of Amsterdam, Postbus 94249, 1090 GE Amsterdam, The Netherlands\\
$^{2}$Instituto de Astronom\'{i}a, Universidad Nacional Aut\'{o}noma de M\'{e}xico, Mexico D.F. 04510, Mexico\\}

\date{Accepted XXX. Received YYY; in original form ZZZ}

\pubyear{2017}

\begin{document}
\defcitealias{parikh2017potential}{Paper~I}

\label{firstpage}
\pagerange{\pageref{firstpage}--\pageref{lastpage}}
\maketitle

\begin{abstract}
We report on two new quiescent {\it XMM-Newton} observations (in addition to the earlier {\it Swift}/XRT and {\it XMM-Newton} coverage) of the cooling neutron star crust in the low-mass X-ray binary 1RXS J180408.9$-$342058. Its crust was heated during the $\sim$4.5 month accretion outburst of the source. From our quiescent observations, fitting the spectra with a neutron star atmosphere model, we found that the crust had cooled from $\sim$ 100 eV to $\sim$73 eV from $\sim$8 days to $\sim$479 days after the end of its outburst. However, during the most recent observation, taken $\sim$860 days after the end of the outburst, we found that the crust appeared not to have cooled further. This suggested that the crust had returned to thermal equilibrium with the neutron star core. We model the quiescent thermal evolution with the theoretical crustal cooling code \nscool and find that the source requires a shallow heat source, in addition to the standard deep crustal heating processes, contributing $\sim$0.9 MeV per accreted nucleon during outburst to explain its observed temperature decay. Our high quality {\it XMM-Newton} data required an additional hard component to adequately fit the spectra. This slightly complicates our interpretation of the quiescent data of 1RXS J180408.9$-$342058. The origin of this component is not fully understood.

\end{abstract}




\section{Introduction}
Matter accreting onto the surface of a neutron star (NS) compresses the underlying layers causing reactions in the crust that release heat. These reactions occur deep in the crust (occurring at $\rho \sim$10$^{12}$ -- $10^{13}$ g cm$^{-3}$) releasing $\sim$1 -- 2 MeV per accreted nucleon of heat disrupting the crust-core thermal equilibrium \citep{haensel1990non,haensel2008models,steiner2012deep}. The most frequently used systems in these studies are NS low-mass X-ray binaries (LMXBs), which consist of a NS and a sub-solar companion that overflows its Roche lobe. The matter transferred in this way will form a disc around the NS and eventually will be accreted onto it. Some systems are `persistent' and the NS is always accreting matter onto its surface from the disc. In other `transient' systems instabilities in the accretion disc \citep[see, for e.g., ][for a review]{lasota2001disc} result only in sporadic episodes of accretion (called outbursts). This accretion heats up the crust. Once the outburst ceases the crust begins to cool in order to reinstate the crust-core equilibrium. Monitoring this crustal cooling (using the observed effective surface temperature) can provide insights into the properties and physics of the high density matter present in the crust \citep[see, for e.g.,][]{shternin2007neutron,brown2009mapping}.

Currently, eight NSs in LMXBs have exhibited crustal cooling when they were monitored after the end of their outbursts \citep[see][for a review]{wijnands2017review}. It has been found that several sources require an extra source of heat (besides the heat released due to the deep crustal reactions) to explain their high observed quiescent temperatures at the earliest phases of the cooling curves (within a few hundreds days after the end of the outbursts). This additional source of heat \citep[typically $\sim$1 -- 2 MeV per accreted nucleon;][]{brown2009mapping,degenaar2014probing,waterhouse2016constraining} is located at a rather shallow depth in the crust of $\rho \sim$10$^{8}$ -- $10^{10}$ g cm$^{-3}$ and therefore is referred to as the `shallow heating' source. The origin of this shallow heating is unknown \citep[see][for a discussion about possible origins]{deibel2015strong}.

1RXS J180408.9$-$342058 (hereafter 1RXS J1804) exhibited a $\sim$4.5 month outburst in early 2015 \citep{barthelmy2015swift,barthelmy2015trigger,krimm2015swift}. In \citet[][hereafter \citetalias{parikh2017potential}]{parikh2017potential}, we have studied the heated NS crust in this system and its thermal evolution upto $\sim$381 days into quiescence, further supporting that outbursts of a few months duration can also heat the NS crust significantly out of equilibrium with the core \citep[see also][]{degenaar2013continued,degenaar2015neutron,waterhouse2016constraining}. To do this, we used several observations obtained using the {\it Swift}/X-ray Telescope (XRT) and one {\it XMM-Newton} pointing. We observed a drop in effective NS surface temperature (\kteff) from $\sim$100 eV to $\sim$71 eV over the probed quiescent period. The {\it XMM-Newton} spectra needed a power-law component (contributing $\sim$30 per cent to the total unabsorbed 0.5 -- 10 keV flux) in addition to the thermal component to model the data well. This additional component was not required by the {\it Swift}/XRT data, although the quality of the {\it Swift}/XRT spectra (low compared to the {\it XMM-Newton} one) was such that a power-law component could not be excluded either. Since the monitoring duration of our observations presented in \citetalias{parikh2017potential} was only about one year, we could only probe the  properties of the shallower crust layers. Observations later in quiescence successively probe deeper layers of the crust. Here we report on two new {\it XMM-Newton} observations that probe the source at a later time after the end of its outburst. We discuss the constraints obtained on the deeper crust properties and remodel the cooling curve in the context of the information provided by these new observations. 

\section{Observations, Data Analysis, and Results}
\begin{figure}
\centering
\includegraphics[scale=0.43]{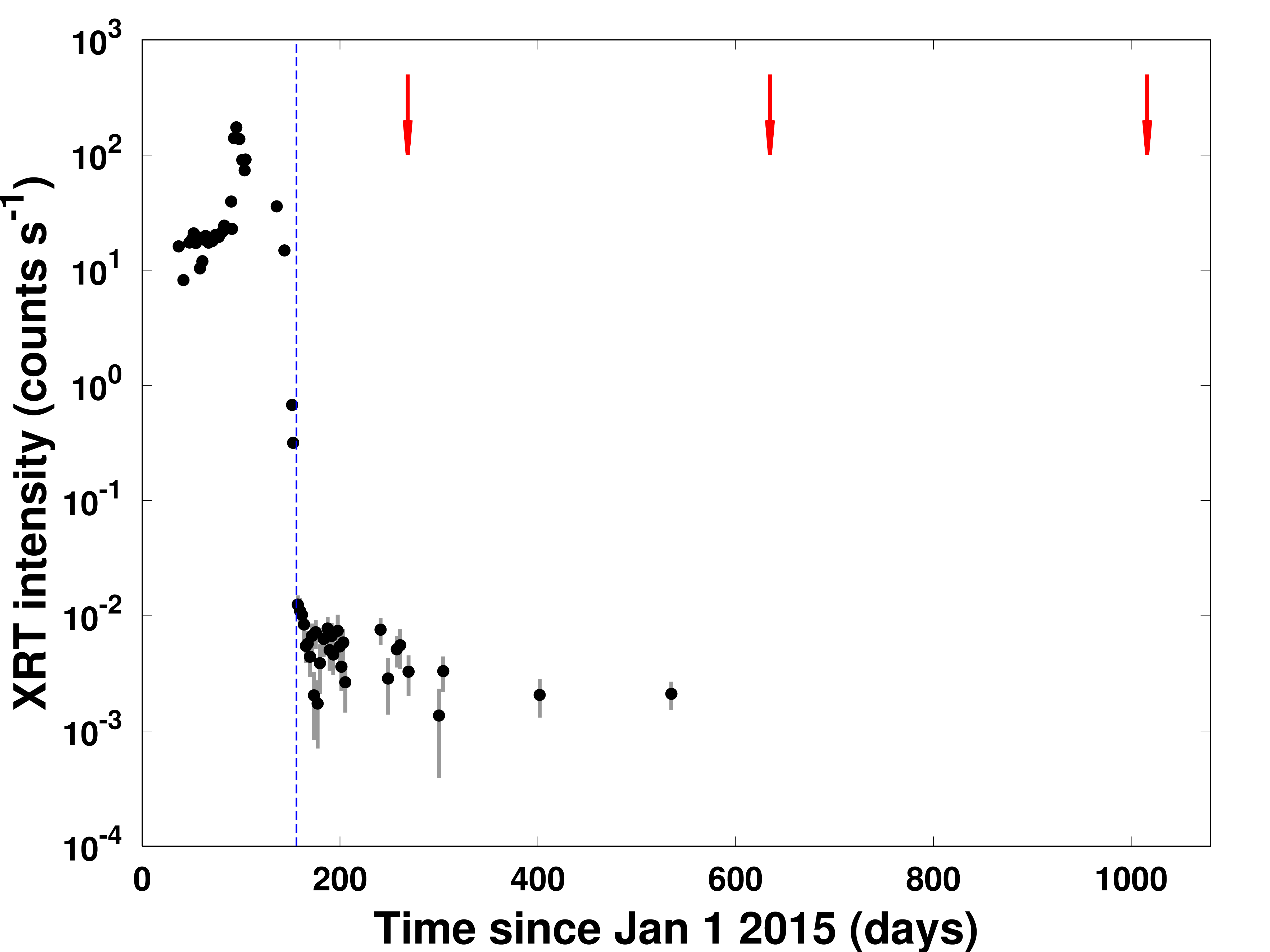}
\caption{The {\it Swift}/XRT (0.5 -- 10 keV) light curve of 1RXS J1804 is shown \citepalias[see][for details about these data]{parikh2017potential}. The dashed blue line indicates the time of transition to quiescence on MJD 57179. The red arrows indicate the times of our three {\it XMM-Newton} observations.}
\label{fig_lc}
\end{figure}

The 2015 outburst of 1RXS J1804 was well covered by the XRT on board the {\it Swift} observatory. The light curve obtained from these observations is shown in Figure \ref{fig_lc}. Several {\it Swift}/XRT observations were also obtained after the source transitioned to quiescence in 2015 June. These quiescent observations were combined into several intervals to determine the early cooling evolution of the crust. As the source cooled further the {\it Swift}/XRT was not sensitive enough to get high quality spectra and we could not use these observations to further constrain the temperature evolution of the source. Therefore the only new observations reported in this work are the two {\it XMM-Newton} observations. More information about the light curve, spectral extraction, and early temperature evolution can be obtained from \citetalias{parikh2017potential} (see also their Table 1 and Table 2).

\subsection{\textbf{\textit{XMM-Newton}}}
{\it XMM-Newton} has observed 1RXS J1804 three times in quiescence at $\sim$112,  $\sim$479, and  $\sim$860 days after the end of its outburst (see Table \ref{tab_info} for a log of the observations; Degenaar and Parikh were the PIs of these observations). During all three observations, the source was observed using all three European Photo Imaging Cameras (EPIC) -- the pn, MOS1, and MOS2 using the full window mode. The results from the first observation (Observation ID [ObsID]: 0770380301) have previously been reported in \citetalias{parikh2017potential}. However, we reanalysed the first observation to ensure a uniform analysis with the most up-to-date software. The data were downloaded from the {\it XMM-Newton} Science Archive\footnote{http://nxsa.esac.esa.int/nxsa-web/\#home} and were reduced using the Science Analysis Software (\texttt{SAS}, version 16.1). The pn and MOS data were processed using \texttt{epproc} and \texttt{emproc}. 

The data were examined for possible background flaring episodes by investigating the light curve between 10 -- 12 keV for the pn data, and $>$10 keV for the MOS data. We found several instances of increased background activity in the three {\it XMM-Newton} observations and removed the data during which the count rate was $>$0.2 -- 0.25 counts s$^{-1}$ for the pn data and $>$0.1 -- 0.7 counts s$^{-1}$ for the MOS data (as was appropriate for the given observation), to eliminate the influence of background flaring. Circular regions were used for the source and background spectral extraction. The source region was determined with the assistance of the \texttt{eregionanalyse} tool which optimises the signal-to-noise ratio. Extraction regions having a radius of 17 -- 21 arcsec were suggested for the pn data and 17 -- 23 arcsec were suggested for the MOS data. The radii suggested depended on the source brightness in the observation concerned. A background region having a radius of 50 arcsec was used throughout. It was placed at a position recommended by the \texttt{ebkgreg} tool, on the same CCD as that the source was located on. The response matrix files and ancillary response functions were created using \texttt{rmfgen} and \texttt{arfgen} respectively. We used \texttt{specgroup} to group our spectra to have a minimum of 25 counts per bin. The details of the three observations, including the exposure times (after the removal of the background flares) and count rates are shown in Table \ref{tab_info}.

\subsection{Spectral Fitting}
We fitted all our {\it XMM-Newton} spectra in the 0.5 -- 10 keV range using $\chi^2$ statistics. The errors presented throughout are given for 90 per cent confidence levels. The spectra were fit using \texttt{XSpec} (version 12.9). We used \texttt{tbabs} to model the equivalent hydrogen column density (\nh), employing \texttt{VERN} abundances and \texttt{WILM} cross-sections \citep{verner1996atomic,wilms2000absorption}. As shown in our previous work \citepalias{parikh2017potential}, 1RXS J1804 hosts a NS that likely exhibits crustal cooling. We used the NS atmosphere model \texttt{nsatmos} \citep{heinke2006hydrogen} to fit the data. We simultaneously fitted all the {\it XMM-Newton} spectra from all three observations of 1RXS J1804 to obtain the best constraints. The NS mass and radius were fixed to $M_\mathrm{NS} = 1.6 \, M_\odot$ and $R_\mathrm{NS} = $ 11 km. We fixed the distance to the source to 5.8 kpc, as determined from the luminosity of its thermonuclear bursts \citep[][assuming an Eddington luminosity limit for helium-rich material]{chenevez2012integral}. We assumed that the entire NS surface was emitting and the normalisation of the \texttt{nsatmos} component was therefore set to 1. The effective temperature was left to vary between different observations but its value was tied together for the pn and MOS data corresponding to a given observation. All the measured effective temperatures were converted to values corresponding to the effective temperature seen by an observer at infinity\footnote{\kteff = $kT_\mathrm{eff}/(1 + z)$, where $(1 + z)$ is the gravitational redshift factor. For $M_\mathrm{NS} = 1.6 \, M_\odot$ and $R_\mathrm{NS} = $ 11 km, $(1 + z)$ = 1.32.} (\kteff).  The \nh value was left free to vary but tied between all the observations as we do not expect it to vary with time. 

Initially, we modelled the spectra using \texttt{nsatmos} and \texttt{tbabs} but we found that the data were not well fitted, having a reduced $\chi^2$ value, $\chi^2_\nu = 1.6$ for 127 degrees of freedom (d.o.f.) This was similar to our results obtained when only fitting the spectra from the first {\it XMM-Newton} observation, as presented in \citetalias{parikh2017potential}. The model needed an additional hard component to fit the spectra well. Therefore, we introduced an additional power-law component -- \texttt{pegpwrlw}, to the model to fit the spectra obtained from all three {\it XMM-Newton} observations. We allowed both the power-law index as well as its normalization to vary freely between the observations. However, once again, these values were tied across the pn and MOS detectors corresponding to a given observation. This improves the fit, $\chi^2_\nu = 1.0$ for 120 d.o.f. We carried out an $F$-test to show that this additional component was statistically required. The probability that this additional component improved the fit by chance was $2.7 \times 10^{-11}$, indicating that this power-law component was necessary to obtain a better fit to our spectra. All three observations, considered individually, similarly indicated the need for this additional component to improve their fit.

The best-fitting \nh was $(0.40 \pm 0.04) \times 10^{22}$ cm$^{-2}$, consistent with that found in our previous study \citepalias{parikh2017potential} as well as those determined by \citet{krimm2015swift2} and \citet{degenaar2016disk}. We fixed the \nh value to 0.4 $\times 10^{22}$ cm$^{-2}$ in order to obtain better constraints on the \kteff and to enable us to directly compare these results to the early \kteff evolution determined from the quiescent {\it Swift}/XRT data \citepalias{parikh2017potential}. A direct comparison was possible as in addition to the \nh value all other \texttt{nsatmos} parameters, such as the NS mass, radius, distance, and surface emission fraction were the same. Furthermore, we have previously shown \citepalias[see Section 3.2 of][]{parikh2017potential} that the \kteff obtained from the {\it Swift}/XRT and {\it XMM-Newton} data could be compared directly even though the {\it XMM-Newton} data needed an additional power-law component to fit its spectra well. The results of the fitting of the {\it XMM-Newton} data are shown in Table \ref{tab_kt}. We calculated the total unabsorbed flux contribution for the 0.5 -- 10 keV energy range using the convolution model \texttt{cflux}. We also calculated the contribution of the power-law component to the total unabsorbed flux (0.5 -- 10 keV). The \kteff determined from the first {\it XMM-Newton} observation is consistent with the value reported for this observation in \citetalias{parikh2017potential}.

\begin{figure}
\centering
\includegraphics[scale=0.34]{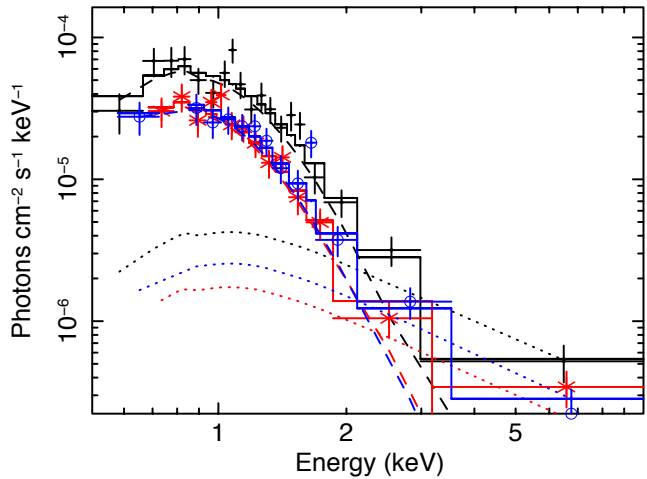}
\caption{The {\it XMM-Newton} X-ray spectra from the MOS2 detector of our three quiescent observations are shown. The black +, red $\ast$, and blue $\circ$ show the data from the first, second, and third observation, respectively. The solid line indicates the best-fitting model, the dashed line indicates the contribution of the \texttt{nsatmos} model, and the dotted line indicates the power-law component (with the index tied across the three observations). }
\label{fig_xmm_spec}
\end{figure}

As can be seen from Table \ref{tab_kt}, the power-law index ($\Gamma$) was not well constrained. The $\Gamma$ was consistent across the three observations within its large error bars. We have summarised the fit results if the $\Gamma$ is allowed to change in Table \ref{tab_kt}. It is not known whether this power-law index should remain the same across different observations or should be allowed to change. We also examine the fit results assuming that the power-law index is the same across the three {\it XMM-Newton} observations. Thus, in our fits we tied the $\Gamma$ value across the various observations. Its normalization was allowed to vary for each observation (but tied between the various {\it XMM-Newton} detectors for a given observation). The best-fitting $\Gamma$ value was $1.6 \pm 0.6$ and the fit indicated a $\chi^2_\nu = 1.0$ for 123 d.o.f. We fixed the power-law index to this best-fit value to obtain better constraints on the \kteff evolution. The spectra from the MOS2 detector, along with the best-fitting models are shown in Figure \ref{fig_xmm_spec} and the \kteff evolution of the source is shown in Figure \ref{fig_nscool}. The \kteff values and the flux contribution of the power-law component to the total 0.5 -- 10 keV flux are shown in Table \ref{tab_kt}. The \kteff and luminosity of the last two {\it XMM-Newton} observations were consistent with one another and indicate that crustal cooling seems to have halted and that the crust in the source has returned to thermal equilibrium with the core. This and other interpretations are further discussed in Section \ref{sect_disc}.

\begin{table}
\centering
\caption{Log of {\it XMM-Newton} observations of 1RXS J1804.\textsuperscript{$a$}}
\label{tab_info}
\begin{tabular}{p{0.17cm}>{\centering\arraybackslash}p{1.1cm}>{\centering\arraybackslash}p{1.1cm}>{\centering\arraybackslash}p{1.75cm}>{\centering\arraybackslash}p{3.4cm}}
\hline
 & ObsID & Date & Exposure& Count Rates  \tabularnewline
 &	           &          & Time (ksec) & ($\times 10^{-3}$ counts s$^{-1}$)\tabularnewline
\hline
 1& 0770380301& 2015/09/26 	 & 6.1, 31.0, 30.8 & 41.2$\pm$2.8, 11.5$\pm$0.7, 14.7$\pm$0.1 \tabularnewline
 2& 0781760101& 2016/09/26  	& 19.5, 72.8, 50.3 & 23.1$\pm$1.2, 5.6$\pm$0.3, 6.4$\pm$0.4 \tabularnewline
3 & 0804910101& 2017/10/13  	& 14.4, 25.9, 41.9 & 22.8$\pm$1.3, 7.7$\pm$0.6, 6.8$\pm$0.4 \tabularnewline

\hline
\multicolumn{5}{p{8.7cm}}{\textsuperscript{$a$}\footnotesize{The exposure times and background subtracted count rates (0.5 -- 10 keV) have been displayed in the order of `pn, MOS1, MOS2'. The exposure times given indicate the effective exposure times obtained after the removal of background flaring.}}\tabularnewline
\end{tabular}
\end{table}

\begin{table*}
\centering
\caption{The spectral fit results obtained for our {\it XMM-Newton} observations of 1RXS J1804.\textsuperscript{$a$}}
\label{tab_kt}
\begin{tabular}{ccccccc}
\hline
 & Days since the& \kteff & $\Gamma$ & $\Gamma$ contribution\textsuperscript{$b$}&Unabsorbed Flux&Unabsorbed Luminosity\tabularnewline
 &end of outburst&(eV) & & (per cent)& ($\times 10^{-13}$ erg cm$^{-2}$ s$^{-1}$)&($\times 10^{32}$ erg s$^{-1}$)\tabularnewline
 \hline
 1 &112.4 &$79.0_{-10.3}^{+3.5}$ & $2.3_{-1.2}^{+0.9}$ & 30.7$_{-12.1}^{+39.0}$ & $2.3\pm0.1$ & $9.1\pm0.5$\tabularnewline
 2 &478.7 & $73.4_{-1.6}^{+1.0}$  & $1.0_{-1.1}^{+1.2}$ & 16.7$_{-5.4}^{+6.1}$ & $1.3\pm0.1$ & $5.3\pm0.3$\tabularnewline
3 &860.2 & $72.0_{-5.0}^{+2.0}$  & $1.7_{-0.9}^{+1.0}$ & 27.6$_{-6.1}^{+15.2}$ & $1.4\pm0.1$ & $5.6\pm0.4$\vspace{0.2cm}  \tabularnewline
\multicolumn{7}{c}{\it $\Gamma$ fixed to the best-fitting value\textsuperscript{$c$}} \vspace{0.2cm} \tabularnewline
 1 &112.4 &$81.2 \pm 1.2$  & 1.6 (fixed) & 23.0$_{-5.6}^{+5.0}$&  $2.3 \pm 0.1$ & $9.3 \pm 0.5$\tabularnewline
 2 &478.7 & $72.8 \pm 1.0$ & 1.6 (fixed) & 16.5$_{-5.2}^{+4.6}$& $1.3 \pm 0.1$ & $5.1 \pm 0.3$\tabularnewline
3 &860.2 &$72.4 \pm 1.3$   & 1.6 (fixed) & 27.0$_{-5.8}^{+5.2}$ & $1.4 \pm 0.1$& $5.6 \pm 0.4$\tabularnewline
\hline
\multicolumn{7}{p{14cm}}{\textsuperscript{$a$}\footnotesize{All errors are reported for the 90 per cent confidence range. The \nh was fixed to 0.4$\times$10$^{22}$cm$^{-2}$ during the spectral fitting. The flux and luminosity (for the 0.5 -- 10 keV energy range) corresponds to the combined power-law and \texttt{nsatmos} flux and luminosities, respectively.} \textsuperscript{$b$}\footnotesize{The $\Gamma$ contribution indicates the flux contribution (in percentage) of the power-law component to the total 0.5 -- 10 keV unabsorbed flux.} \textsuperscript{$c$}\footnotesize{The $\Gamma$ was fixed to the value obtained when fitting the spectra with $\Gamma$ tied between all data sets.}}\tabularnewline
\end{tabular}
\end{table*}

\subsection{Modelling the observed \kteff evolution}
\label{sect_nscool}
We modelled the quiescent \kteff evolution of 1RXS J1804, observed using {\it Swift}/XRT and {\it XMM-Newton}, using the crustal cooling code \nscool \citep{page2013forecasting,page2016nscool}. We accounted for the variability in accretion rate during the outburst \citep[][using daily averaged accretion rates]{ootes2016}. We determined this variability using the outburst light curves from the {\it Swift}/XRT and {\it MAXI} instruments. The details are provided in \citetalias{parikh2017potential}. In our modeling, we chose the \kteff data obtained by fixing the power-law index to the best-fitting value to model the cooling evolution since we do not know how the power-law index should behave. Leaving the index free to vary between observations resulted in weaker constraints from the \texttt{nsatmos} model. The \kteff evolution determined by fixing the power-law index to its best-fitting value is more constraining and consistent with the \kteff estimates obtained when the power-law index is allowed to vary.

\begin{figure}
\centering
\includegraphics[scale=0.55]{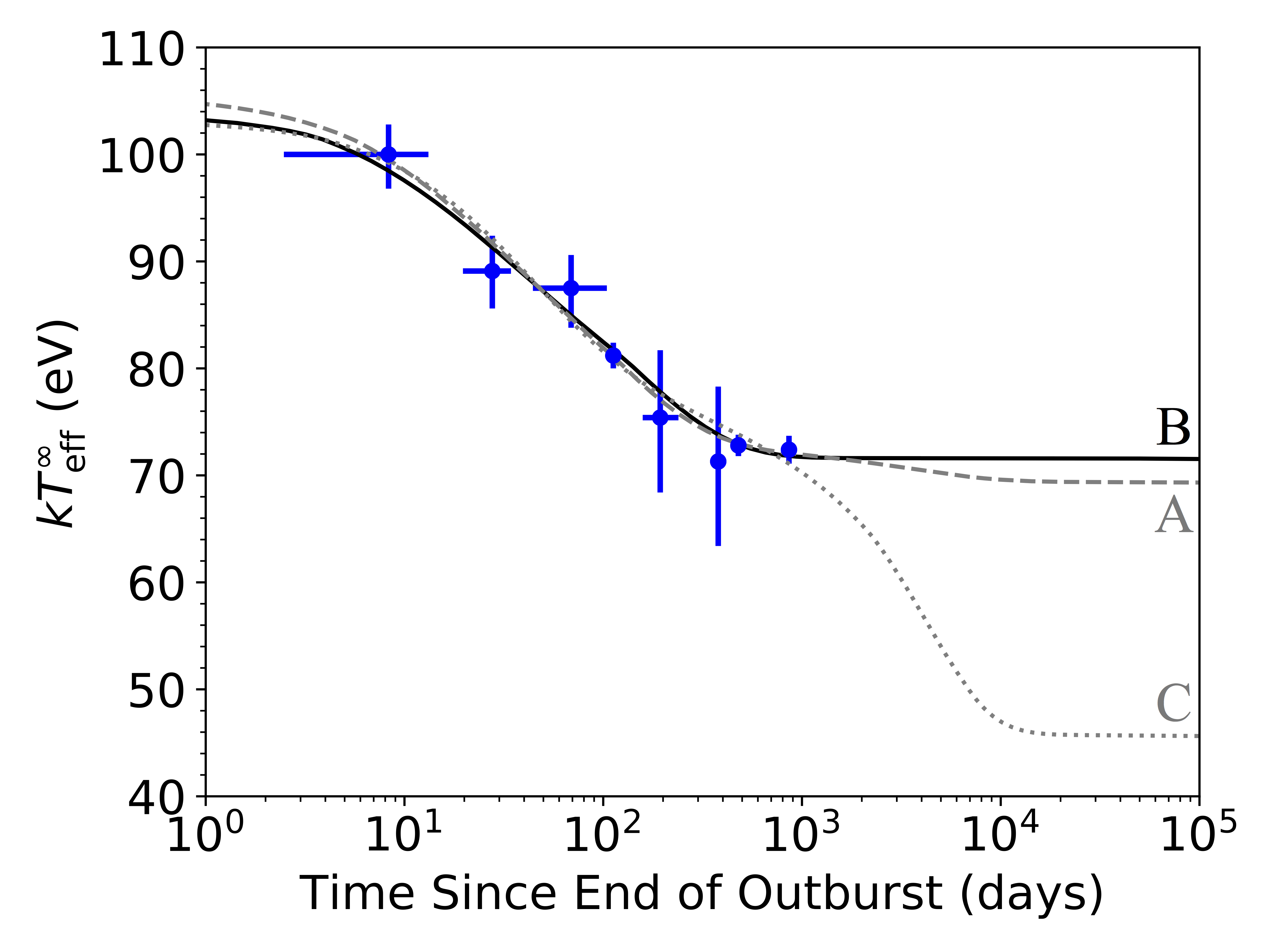}
\caption{The observed thermal evolution of 1RXS J1804 (blue data points) as well as some of our \nscool fits through these data points are shown. Model A (dashed grey line) indicates the model if the $Q_\mathrm{imp}$ is free to vary. The $Q_\mathrm{imp}$ was found to be unconstrained, therefore, we fixed the $Q_\mathrm{imp}$ = 1 throughout the crust and the corresponding model is Model B (solid black line). Including high impurity zones in the crust potentially allows for a very low base level. Model C (dotted grey line) shows the minimum core temperature that can be obtained within the 90$\%$ confidence level of Model A. A summary of these parameters is shown in Table \ref{tab_nscool_val}.}
\label{fig_nscool}
\end{figure}

We fixed the NS mass and radius in the \nscool model to the same values as those used during the spectral fitting ($M_\mathrm{NS} = 1.6 \, M_\odot$ and $R_\mathrm{NS} = $ 11 km). Our model assumes that 1.93 MeV per accreted nucleon of deep crustal heating is active \citep{haensel2008models,ootes2016}. The remaining parameters were left free to vary. These parameters are the core temperature ($T_0$), the strength and depth of the shallow heating source ($Q_\mathrm{sh}$ and $\rho_\mathrm{sh}$), the composition of the light elements in the envelope ($y_\mathrm{light}$), and the impurity factor in the crust ($Q_\mathrm{imp}$).  This $Q_\mathrm{imp}$ is modelled as three layers.\footnote{The upper and lower boundary of the crust is defined by $\rho$ = $10^{8}$ g cm$^{-3}$ and $\rho$ = $1.5 \times 10^{14}$ g cm$^{-3}$, respectively. The two intermediate density boundaries that define the three crustal layers (which can be modelled using different $Q_\mathrm{imp}$) are $\rho$ = $4 \times 10^{11}$ g cm$^{-3}$ and $\rho$ = $8 \times 10^{13}$ g cm$^{-3}$. \label{footnote_rho}} The lowest layer contains the pasta phase \citep{horowitz2015disordered} and extends to the crust-core boundary (see footnote \ref{footnote_rho} for details about the density over which this pasta layer extends). Leaving the $Q_\mathrm{imp}$ free to vary resulted in it being unconstrained in all layers of the crust (because we do not allow it to go below zero or above 300). The model with these parameters is shown as Model A (indicated by the dashed grey line) in Figure \ref{fig_nscool}. 

Since the $Q_\mathrm{imp}$ was completely unconstrained, we fixed it to 1 throughout the crust.  Several NS crusts in LMXBs are consistent with having a low impurity crust corresponding to $Q_\mathrm{imp}$ = 1 \citep[e.g., Swift J174805.3$-$244637, Aql X$-$1,  MAXI J0556-332;][]{degenaar2015neutron,waterhouse2016constraining,parikh2017different} although a few sources also have $Q_\mathrm{imp} >$ 1 \citep[e.g., EXO 0748$-$676, KS 1731$-$261;][]{degenaar2014probing,ootes2016,merritt2016neutron}. The best-fitting \nscool model, with $Q_\mathrm{imp}$ fixed to 1, indicated as Model B (shown by the solid black line) in Figure \ref{fig_nscool} and has $\chi^2_\nu = 0.5$ for 4 d.o.f. The errors on the model parameters are given for the 90 per cent confidence level. The best fit for Model B indicates a core temperature of $T_0 = 4.7_{-0.8}^{+4.1} \times 10^{7}$ K. A summary of the $T_0$ and $Q_\mathrm{imp}$ parameters for the various models are shown in Table \ref{tab_nscool_val}. For Model B we obtained a best-fitting $y_\mathrm{light}$ = $1.8 \times 10^{9}$ g cm$^{-2}$ with the errors pegging at the lower and upper boundary. The lower and upper boundary (corresponding to $y_\mathrm{light}$ = $10^{3}$ g cm$^{-2}$ and $10^{11}$ g cm$^{-2}$, respectively) are indicative of the limits of the composition of light elements in the envelope beyond which there is no effect of the change in their composition. The shallow heating parameters correspond to $Q_\mathrm{sh}$ = $0.9_{-0.2}^{+0.8}$ MeV per accreted nucleon at a depth $\rho_\mathrm{sh}$ = $2.9_{\ast}^{+42.3} \times 10^{8}$ g cm$^{-3}$ (where $\ast$ indicates that the fit has pegged to the lowest $\rho_\mathrm{sh}$ of $10^{8}$ g cm$^{-3}$ which corresponds to the NS crust-surface boundary). These $Q_\mathrm{sh}$ and $\rho_\mathrm{sh}$ values are consistent with those reported in \citetalias{parikh2017potential}. Furthermore, they are also consistent with the values for these parameters obtained using Model A, where the $Q_\mathrm{imp}$ is free to vary.

\begin{table}
\centering 
\caption{Parameter values of the three \nscool models.}
\label{tab_nscool_val}
\begin{tabular}{llp{5.5cm}}
\hline
Model & $T_0\, (\times 10^{7}$ K)& Notes on the $Q_\mathrm{imp}$ \tabularnewline
\hline
A & 5.7$_{-3.9}^{+3.0}$ & $Q_\mathrm{imp}$ unconstrained in all three layers \tabularnewline
B & 4.7$_{-0.8}^{+4.1}$ & $Q_\mathrm{imp}$ fixed to 1 in all three layers \tabularnewline
C & 1.8 (fixed) & $Q_\mathrm{imp}$ = 1 in the shallowest layer, $Q_\mathrm{imp}$ = 55 in the middle layer, and $Q_\mathrm{imp,pasta}$ pegs at the highest allowed value of 300 \tabularnewline
\hline
\end{tabular}
\end{table}



\section{Discussion}
\label{sect_disc}
We present the results of our two new {\it XMM-Newton} observations of the quiescent NS LMXB 1RXS J1804. Previous quiescent observations of the source (presented in \citetalias{parikh2017potential}) indicated that it hosted a NS whose crust was significantly heated (disrupting the thermal equilibrium with the core) during the preceding $\sim$4.5 month outburst and subsequently exhibited cooling during quiescence. 

Fitting a \texttt{nsatmos} model to the spectra obtained from our observations, we found that the effective surface temperature of the source had dropped from \kteff $\sim$100 eV to $\sim$73 eV, over $\sim$479 days. However, it appeared not to have dropped further after that because $\sim$860 days after the end of the outburst the \kteff was approximately the same. The consistent \kteff observed during our last two {\it XMM-Newton} observations suggested that the NS crust was in thermal equilibrium with the core. This was also inferred from the best-fitting model calculated using our cooling code \nscool. To explain the observed behaviour, our model required $\sim$0.9 MeV per accreted nucleon of shallow heating to be active during the outburst, in addition to the standard deep crustal heating. 

The plateauing behaviour exhibited by the \kteff evolution of 1RXS J1804 during the last two observations has previously been observed for several other crustal cooling sources \citep[see Figure 2 of][for the crustal cooling curves of all sources studied so far]{wijnands2017review}. This was interpreted by several authors to indicate that the crust and core were (almost) in equilibrium again \citep{fridriksson2010rapid,degenaar2011further}. However, subsequent observations of some these sources, such as XTE J1701$-$462, EXO 0748$-$676, and KS 1731$-$260, showed that the effective surface temperature dropped further demonstrating that equilibrium was not yet attained. The plateauing behaviour for these sources occurred around similar times after the end of their respective accretion outbursts \citep[$\sim$500 -- 1000 days after the end of the outburst for XTE J1701$-$462 and EXO 0748$-$676, and a little bit later, at $\sim$800 -- 1500 days, for KS 1731$-$260; ][]{fridriksson2011variable,degenaar2014probing,merritt2016neutron}. The sources have limited observational coverage in this phase, which inhibits us from making strong conclusions. However, it is interesting to note that several sources have now shown similar behaviour. Our source 1RXS J1804 also exhibits its plateau phase at a time similar to these other sources after the end of its outburst. The cause for this apparent plateau phase is not known. 

One of the hypotheses for such a plateau to occur is that it may be a result of compositionally driven convection due to the chemical separation of the light and heavy elements when the liquid ocean cools and begins to solidify \citep{horowitz2007phase,medin2011compositionally}. For XTE J1701$-$462, \citet{medin2014signature} modelled the cooling curve with this chemical separation which showed a plateau similar to what we observe for 1RXS J1804. This plateau in XTE J1701$-$462 is followed by further cooling. \citet{degenaar2014probing} show that this chemical separation may also help explain the plateau phase in EXO 0748$-$676.

Alternatively, this plateau phase may be a result of the low thermal conductivity due to the disordered nuclei expected in the pasta layer \citep[present at $\rho \sim 10^{14}$ g cm$^{-3}$;][]{horowitz2015disordered}. \citet{merritt2016neutron} attempted to include such a low conductivity pasta layer in their model to fit the observed cooling curve of KS 1731$-$260. Their findings were inconclusive as they obtained models that could describe the data both with and without the inclusion of this disordered pasta layer. \citet{deibel2017late} also examined the cooling curve of KS 1731$-$260 as well as that of MXB 1659$-$29. For both sources, they found that the fits using their theoretical model preferred a low conductivity pasta layer that showed late-time cooling.

In order to investigate whether 1RXS J1804 could cool down further due to the potential presence of a disordered pasta layer, we fixed the core temperature to the lowest bound permitted by the errors on our best-fitting model when the $Q_\mathrm{imp}$ was also free to vary (Model A; see also Section \ref{sect_nscool} and Table \ref{tab_nscool_val}). The core temperature was fixed to $T_0 = 1.8 \times 10^{7}$ K and the cooling evolution was modelled using the observed quiescent \kteff data. This model is indicated by Model C (shown by the dotted grey line) in Figure \ref{fig_nscool}. We find that Model C shows a plateau phase, followed by a further drop in \kteff in the future. The plateauing behaviour predicted by this model is indeed a result of the low conductivity of the disordered nuclei expected at this depth in the crust. The $Q_\mathrm{imp}$ in this layer (the $Q_\mathrm{imp,pasta}$) is very high and pegs at the largest $Q_\mathrm{imp,pasta}$ allowed in our model, at $Q_\mathrm{imp,pasta} =$ 300 (with the $Q_\mathrm{imp}$ remaining 1 for the outer crust and $Q_\mathrm{imp}$ = 55 for the middle crust layer; see also Table \ref{tab_nscool_val} and footnote \ref{footnote_rho}). This $Q_\mathrm{imp}$ value is much larger than the $Q_\mathrm{imp,pasta}$ of $\sim$20 -- 40 determined by \citet{horowitz2015disordered} from the electron-pasta scattering based on their predicted disordered nuclei structure. However, we present Model C as an extreme case (having a low $T_0$) to show the largest amount of late-time cooling that might still be possible in our modelling, although the physical validity of this model remains to be determined. The drop in \kteff after the plateau phase is caused by accelerated cooling once the heat from the deeper crustal layers, beneath those hosting the low thermal conductivity pasta, begins to propagate outwards. 

We find that 1RXS J1804 can only show this post-plateau phase drop if the $Q_\mathrm{imp,pasta}$ is high {\it and} the core temperature is low. If the core temperature is higher, as indicated by Model A, the $Q_\mathrm{imp,pasta}$ is unconstrained and the model can still accommodate a $Q_\mathrm{imp,pasta} =$ 300 without showing this subsequent drop. This would suggest that the actual drops that have been observed for several other sources are likely due to the presence of the low conductivity pasta layer in combination with a large temperature gradient (i.e. significant heating causing a strong thermal gradient in the crust with respect to the core temperature). We know that significant accretion-induced crustal heating is possible for the three sources that show this drop since they had long outbursts \citep[an exceptionally bright $\sim$1.5 year outburst of XTE J1701$-$462, $\sim$24 year and $\sim$12.5 year outbursts of EXO 0748$-$676 and KS 1731$-$260, respectively; ][]{wijnands2001chandra,degenaar2009chandra,fridriksson2010rapid}. However, 1RXS J1804 only had an intermediately long outburst of $\sim$4.5 months \citepalias[see][]{parikh2017potential} and it is unknown if it has a strong enough thermal gradient in the deep crust with respect to its core temperature. We need observations $\sim$1000 days (or further) after our last {\it XMM-Newton} observation in order to investigate whether or not 1RXS J1804 will cool further.

In \citetalias{parikh2017potential}, we showed that no deep crustal heating was necessary to explain the \kteff evolution that was observed at that time. The new observations we present in this paper probe deeper layers of the NS crust. We repeated the test performed in \citetalias{parikh2017potential} to determine if our observed cooling curve, including the late time observations, could still be explained with no deep crustal heating. Once again, we found that this deep crustal heating does not need to be invoked to explain the \kteff decay observed for 1RXS J1804. The strength of the shallow heating for this model is $Q_\mathrm{sh}$ = $1.5_{-0.3}^{+0.7}$ MeV per accreted nucleon acting at a depth $\rho_\mathrm{sh}$ = $4.4_{\ast}^{+45.7} \times 10^{8}$ g cm$^{-3}$(where $\ast$ indicates that the fit has pegged to the lowest $\rho_\mathrm{sh}$ of $10^{8}$ g cm$^{-3}$ which corresponds to the NS crust-surface boundary). These parameter values are consistent with the ones where deep crustal heating was also active in our model (see Section \ref{sect_nscool}). Although the presence of the deep crustal heating mechanism is expected in the heating models, it remains to be determined if this process is indeed active in this source. Moreover, similar investigations of other sources have to be performed to determine if the deep crustal heating process is needed at all to explain the available cooling curves.

\subsubsection*{Inferring properties of the NS core}
\begin{figure}
\centering
\includegraphics[scale=0.7]{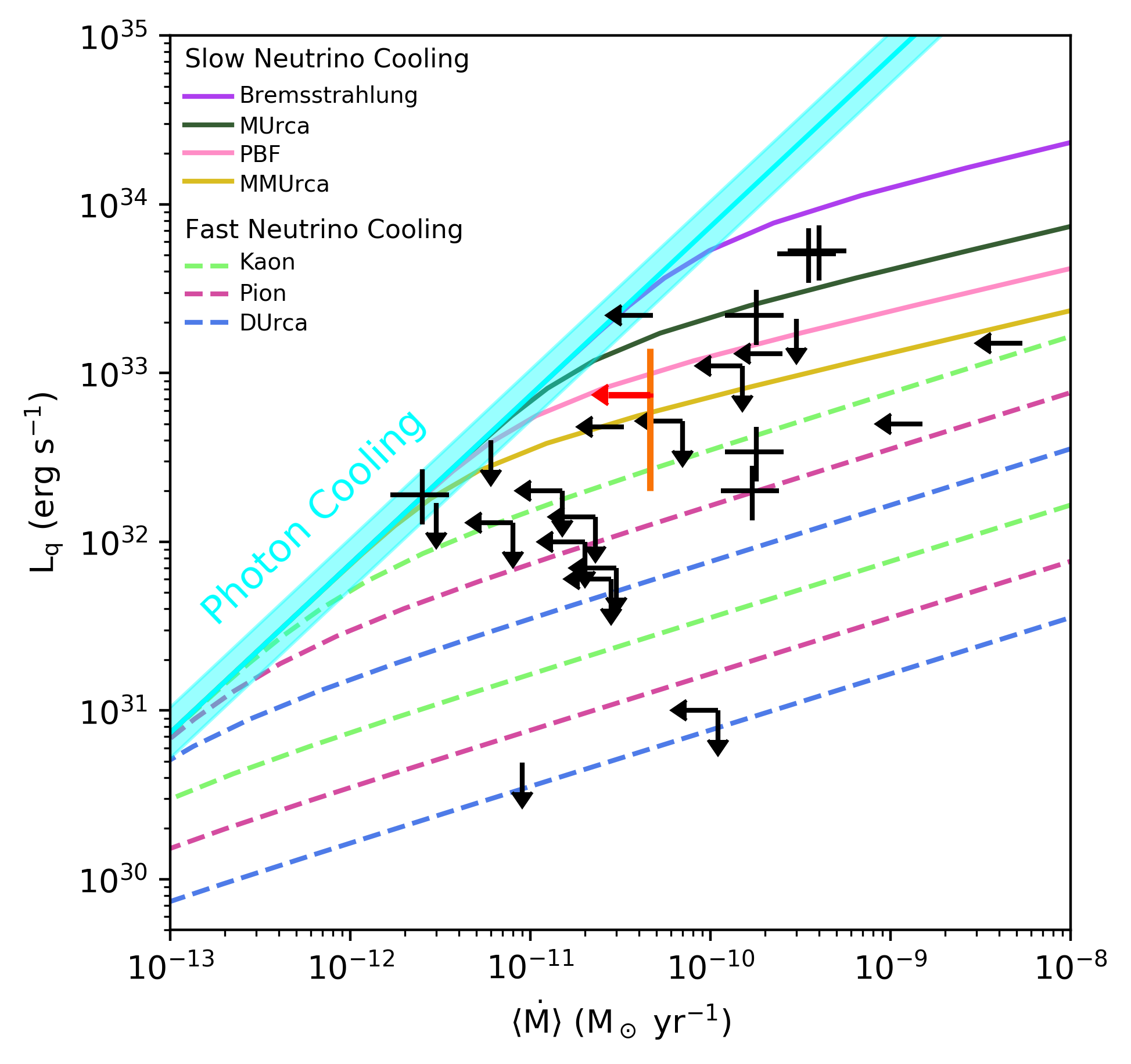}
\caption{The $L_\mathrm{q}$ versus \mdot data from several NS sources in quiescence are shown in black. These data have been obtained from Table 2 of  \citeauthor{heinke2007constraints} (\citeyear{heinke2007constraints}; in addition, see \citeauthor{heinke2010discovery} \citeyear{heinke2010discovery}, for the NGC 6440 X-2 data). The result corresponding to 1RXS J1804 is shown in red. The solid orange line indicates the error on the $L_\mathrm{q}$ estimate of 1RXS J1804 if it has extreme envelope compositions, consisting of mostly light or mostly heavy elements. In addition, we also show theoretical curves for the various cooling mechanism that could be at work in the NS core \citep[see][for more information on the observational data of other sources presented in this plot and details about the cooling curves]{wijnands2013testing}.}
\label{fig_lxvmdot}
\end{figure}

If the current quiescent crustal temperature of 1RXS J1804 is representative of the core temperature it can help us infer properties of the NS core. Comparing the quiescent X-ray luminosity ($L_\mathrm{q}$) with the estimated time-averaged mass accretion rate \mdot onto 1RXS J1804 can indicate the possible core cooling mechanisms at work and from that an indication of the mass of the NS \citep[e.g.,][]{colpi2001charting}. This has previously been done for various quiescent NS LMXB sources \citep[e.g., see][]{yakovlev2004neutron,heinke2007constraints,heinke2009further,heinke2010discovery, wijnands2013testing}. The $L_\mathrm{q}$ corresponds to the bolometric luminosity that quantifies the internal heat from the NS (when the crust is in thermal equilibrium with the core). We needed a soft as well as hard component to model our spectra. We assume that only the soft component\footnote{Although we have used only the soft component of the spectra to determine the $L_\mathrm{q}$, previous studies of this type (including the various sources shown in Fig. \ref{fig_lxvmdot}) may have been determined by using the total contributing luminosity from the both the soft as well as any possible hard spectral component, if present. Thus it is likely that the $L_\mathrm{q}$ from some of those sources was overestimated and actually a lower value needs to be used. However, it is beyond the scope of the current paper to investigate all the previous $L_\mathrm{q}$ determinations. There may be similar uncertainties in the \mdot estimation of the various sources shown in Figure \ref{fig_lxvmdot} as they have been calculated by different authours using different assumptions. Our conclusions applicable to 1RXS J1804 are still valid as we compare our observational data to the theoretical cooling curves and they are independent of the data from these other sources.} is representative of the $L_\mathrm{q}$. The unabsorbed 0.5 -- 10 keV luminosity from the soft component of the last two observations in quiescence was $L_\mathrm{X,soft} = (4.3 \pm 0.3) \times 10^{32}$ erg s$^{-1}$ and $L_\mathrm{X,soft} = (4.1 \pm 0.3) \times 10^{32}$ erg s$^{-1}$, respectively, and therefore we assume an averaged $L_\mathrm{X_{avg},soft} = (4.2 \pm 0.2) \times 10^{32}$ erg s$^{-1}$. This corresponds to a 0.01 -- 100 keV bolometric luminosity of $L_\mathrm{q} \sim 7.4 \times 10^{32}$ erg s$^{-1}$.

The average mass accretion rate onto the NS during the 2015 outburst, taking into account the variability in the accretion rate, was determined to be $\sim$9.3 $\times 10^{-10}$ M$_\odot$ yr$^{-1}$ \citepalias[see][]{parikh2017potential}. This has been the only known significant outburst of 1RXS J1804. In the last two decades the X-ray sky has been monitored in depth with existing all-sky monitors. We assume that if 1RXS J1804 had shown significant activity, prior to its 2015 outburst, it would have been detected by such all-sky monitors. 1RXS J1804 exhibited some low-level activity in 2011 and 2012 (see Section \ref{sect_disc_powerlaw}). However, this activity is not significant enough to change our \mdot value and we do not take it into account. Other similar activity at such a low level would not be detected by the all-sky monitors so it is likely that we have missed these. However, we have extensively monitored 1RXS J1804 up to $\sim$860 days after the end of its outburst and have observed no such type of activity. Thus, we assume that such low-level activity does not occur very frequently. Occasionally occurring low-level activity does not significantly alter our estimated \mdot and therefore we do not take it into account. It is possible that 1RXS J1804 has experienced some significant activity during its Sun constraint windows. We cannot further constrain such a possibility and therefore, we assume that there was no significant activity in any such window. Other outbursts, having a brightness and length similar to the 2015 outburst (i.e., significantly longer than the duration of the Sun constraint windows and detectable by the all-sky monitors) of 1RXS J1804, have not been detected and therefore we assume a quiescent period of at least 20 years in our determination of the \mdot for this source. We calculated the time averaged mass accretion rate for 1RXS J1804 to be \mdot $<$ 4.6 $\times 10^{-11}$ M$_\odot$ yr$^{-1}$. 

We use our estimated \mdot and $L_\mathrm{q}$ for 1RXS J1804, as shown by the red data in Figure \ref{fig_lxvmdot}, and compare it to the various theoretical cooling models. The data point corresponding to 1RXS J1804 lies in between the (standard) slow neutrino cooling curves. Thus, 1RXS J1804 likely does not host a massive NS \citep[determined using the cooling mechanism as an indication for the NS mass, see][]{colpi2001charting}. However, there are many uncertainties on both its $L_\mathrm{q}$ and \mdot. Our assumption that 1RXS J1804 has not exhibited any significant activity in the last $\sim$20 yr, apart from the 2015 outburst, is limited by the relatively recent existence of all-sky monitors. It could also be that this source has not shown a significant outburst for a longer time and thus our \mdot assumption is only indicative of an upper limit. This is represented by the leftward pointing arrow for 1RXS J1804 in Figure \ref{fig_lxvmdot}. The light blue line in Figure \ref{fig_lxvmdot} indicates the regime defined only by photon cooling i.e. not requiring any neutrino cooling mechanism \citep{page2004minimal,wijnands2013testing}. If 1RXS J1804 did not show a significantly strong outburst (apart from that in 2015) for the last $\sim$100 years it would be consistent with the photon cooling regime. This could be possible for 1RXS J1804 within our limited knowledge of its \mdot. Alternatively, it is also possible that during the last 20 years the source exhibited a lull in its activity and that typically during the last tens of thousands of years the source was significantly more active (e.g., an outburst once every several years). If this is the case, the source would move to the right in Figure \ref{fig_lxvmdot} and neutrino emission mechanisms should then definitely be taken into account.

If 1RXS J1804 continues to cool, then the source would move to lower $L_\mathrm{q}$ in Figure \ref{fig_lxvmdot} and enhanced core cooling by fast neutrino processes might be needed indicating that our source may host a massive neutron star. Future observations will be able to conclusively confirm or reject this possibility.

An additional uncertainty on the $L_\mathrm{q}$ is the composition of the envelope (that constitutes the outer $\sim$100 m of the crust). The inferred NS temperature seen by an observer depends on the envelope as it translates the temperature of the NS crust to the surface \citep[e.g.,][]{potekhin1997internal}. The best-fitting envelope composition determined using the \nscool model is shown in Section \ref{sect_nscool}. However, if the envelope composition was different then we would have inferred a different $L_\mathrm{q}$ \citep[see also the study by][]{han2017cooling}. This changing envelope composition does not affect our current study since the envelope composition is not expected to change during quiescence (unless significant low-level accretion occurs). However, if 1RXS J1804 experiences another outburst this envelope composition may change which would result in a change in our inferred $L_\mathrm{q}$ even if the core temperature remains the same \citep{brown2002variability}. To illustrate this effect we varied the composition of the envelope to the two most extreme cases: a very light element envelope (defined by $y_\mathrm{light}$ = $10^{11}$ g cm$^{-2}$) and a very heavy element envelope (defined by $y_\mathrm{light}$ = $10^{3}$ g cm$^{-2}$), assuming the same core temperature as inferred from our crustal cooling study (also indicative of the crust since it is assumed to be in equilibrium with the core). Such extreme envelopes would give rise to a surface bolometric luminosity of $\sim 14 \times 10^{32}$ erg s$^{-1}$ and $\sim 2 \times 10^{32}$ erg s$^{-1}$, respectively, as indicated by the orange error bars in Figure \ref{fig_lxvmdot}. The envelope composition is difficult to infer. If we do not correctly account for it and model 1RXS J1804 using a very light envelope we may interpret the source as not needing any fast neutrino cooling mechanism. Alternatively, if we modelled it using a heavy element envelope we would interpret it as definitely requiring fast neutrino cooling. This highlights the uncertainties the chemical composition can introduce in these studies and this could affect most, if not all, of the source presented in Figure \ref{fig_lxvmdot} \citep[see also][]{han2017cooling}.

\subsection{The additional hard component}
\label{sect_disc_powerlaw}

The quiescent \kteff evolution of 1RXS J1804 indicates a cooling accretion-heated NS crust. However, our {\it XMM-Newton} observations show the need for an additional hard component to describe the spectra well. This complicates our interpretation. 

We have modelled this additional contribution using a power-law and found that it cannot be well constrained. In the case where the power-law index was left free to vary between observations we find that the indices were consistent within their errors as these errors were large (around $\pm 1$). Leaving the indices free was not constraining; tieing them between observations improved the constraints with $\Gamma = 1.6 \pm 0.6$, however $\Gamma$ could still be quite hard or quite soft within its error bars. This was also reflected in the contribution of this power-law component to the total unabsorbed flux which could be any value between $\sim 10 - 30$ per cent if the power-law index was tied between the various observations, and even up to $\sim$ 70 per cent if the power-law index was free to vary (albeit with very large errors on this contribution). Therefore, it is difficult to determine the exact contribution of the power-law component in the spectra. In order to obtain stronger constraints from this power-law component we need to make more assumptions (such as the  variability of the power-law index and its contribution to the total flux) which may or may not be valid. Therefore, we have not investigated this further.

The fractional contribution of the power-law component to the total flux in 1RXS J1804 is consistent across the three epochs we have observed (see Table \ref{tab_kt}). This is different from the behaviour of power-law component in the spectra of XTE J1701$-$462 and EXO 0748$-$676 which varied non-monotonically with time \citep{fridriksson2011variable,degenaar2014probing}. 
The origin of this power-law component in quiescent NS spectra is not fully understood. One of the hypotheses is that it arises from low-level accretion onto the NS  \citep[see e.g.,][]{rutledge2002variable,cackett2010quiescent,chakrabarty2014hard,d2015radiative,wijnands2015low}. 

Prior to the 2015 outburst 1RXS J1804 was observed in 2011 using {\it Chandra}. The source was observed at a $L_\mathrm{X} \sim 2.2 \times 10^{33}$ erg s$^{-1}$, which was significantly higher than our highest observed $L_\mathrm{X}$ after the end of the 2015 outburst \citepalias[see][]{parikh2017potential}. Thus, this pointing was likely obtained when the source exhibited some low-level activity. This rather high value of $L_\mathrm{X}$ is unlikely to be from an accretion-heated crust since no strong outburst was observed preceding this observation. Studying the spectrum indicated the need for a power-law component, in addition to the soft component. However, the parameters of this power-law component could not be constrained due to the limitations of the data quality. Thus, this {\it Chandra} observation shows that 1RXS J1804 has previously experienced some elevated activity during quiescence likely due to low-level accretion. In addition to this {\it Chandra} observation, \citet{chenevez2012integral} reported a thermonuclear type-I burst from this source in early 2012. Follow-up observations carried out within a day indicated that the 1RXS J1804 was already at a low luminosity of $L_\mathrm{X} \sim 3 \times 10^{33}$ erg s$^{-1}$ \citep{kaur2012swift}. It is unknown if this was also indicative of some low-level activity during quiescence or was the end of a small outburst that was not detected by the all-sky monitors. Thus, there is evidence suggesting that 1RXS J1804 experienced some low-level accretion activity in the past. 

Our quiescent observations (after the end of the 2015 outburst) indicate a smooth decay whereas low-level accretion is expected to show variability (as sudden increases in activity in the form of flares and/or more stochastic increases or decreases in activity). However, the behaviour of this low-level accretion is not understood and it could also exhibit a smooth monotonic decay such as that we have observed. A smooth \kteff evolution profile as a result of low-level accretion has been reported for the black hole LMXB Swift J1357.2$-$0933 \citep{armas2013xmm}. However, this smooth decay lasted for a relatively short period of $\sim$100 days whereas we have observed a smooth decay for 1RXS J1804 for $\sim$900 days. It is unknown if low-level activity would exhibit a smooth decay for such a long time but it is not expected. Therefore, we assume that 1RXS J1804 indeed hosts a NS crust that exhibits cooling. This is further supported by the results from our theoretical crust heating/cooling code \nscool whose best-fitting model indicates a NS crust that cools in quiescence.

\section*{Acknowledgements}

AP, RW, and LO are supported by a NWO Top Grant, Module 1, awarded to RW. ND is supported by an NWO Vidi grant. DP is partially supported by the Consejo Nacional de Ciencia y Tecnolog{\'\i}a with a CB-2014-1 grant $\#$240512.




\bibliographystyle{mnras}
 \newcommand{\noop}[1]{}



%
%


\bsp	
\label{lastpage}
\end{document}